\documentstyle[12pt,preprint,aps]{revtex}

\newcommand{\be}{\begin{equation}}
\newcommand{\ee}{\end{equation}}
\newcommand{\ba}{\begin{eqnarray}}
\newcommand{\ea}{\end{eqnarray}}
\newcommand{\bd}{\begin{displaymath}}
\newcommand{\ed}{\end{displaymath}}

\def\thalf{{\textstyle{\frac{1}{2}}}}

\title{Neutrino Superfluidity}

\author{J. I. Kapusta\footnote{kapusta@physics.umn.edu}}
\address{
School of Physics and Astronomy\\
University of Minnesota\\
Minneapolis, MN 55455}

\date{12 November 2004}

\begin{document}

\maketitle
\begin{abstract}

It is shown that Dirac-type neutrinos display BCS superfluidity for 
any nonzero mass.  The Cooper pairs are formed by attractive scalar Higgs boson 
exchange between left and right-handed neutrinos; in the standard $SU(2) \times 
U(1)$ theory, right-handed neutrinos do not couple to any other boson.  The 
value of the gap, the critical temperature, and the Pippard coherence length are 
calculated for arbitrary values of the neutrino mass and chemical potential.  Although such a superfluid could conceivably exit, detecting it would be a major challenge.

\end{abstract}

\vspace*{0.5in}

\newpage

Superconductivity is ubiquitous in nature.  It occurs in metals, organic 
compounds, atomic and molecular gases, nuclear matter, and quark matter.  It 
will be demonstrated here that massive Dirac-type neutrinos can display 
superfluidity, the analogue of superconductivity for electrically neutral 
particles, when they are embedded in the standard model of particle physics.  
What allows neutrino superfluidity is the attractive scalar interaction between 
left and right-handed neutrinos (assuming that right-handed neutrinos exist) due 
to their coupling to the Higgs field from which they obtain their common mass.  As was shown by Caldi and Chodos \cite{Caldi}, pairing of left-handed neutrinos with left-handed neutrinos cannot occur due to the repulsive vector interaction 
provided by the Z$^0$ meson.  For the purpose of demonstration, in this paper 
only one flavor of lepton is considered.  More flavors open the window to 
neutrino oscillations and an even richer structure of superfluid states.  The 
word neutrino is used in a generic sense to denote any electrically neutral 
spin-$\thalf$ point-like fermion.

In a convenient gauge the Higgs field can be written as
\be
\Phi = \frac{1}{\sqrt{2}} \left( \begin{array}{c}
0 \\ v_0 + \sigma \end{array} \right)
\ee
where $v_0 = 1/\sqrt{\sqrt{2}G_{\rm F}} = 246$ GeV is the Higgs condensate and 
$\sigma$ is its fluctuation.  A Yukawa coupling between the neutrino and the 
Higgs field gives the neutrino its mass of $m_{\nu} = h_{\nu} v_0/\sqrt{2}$, 
namely,
\be
{\cal L}_{\rm Y} = h_{\nu} \bar{l}_{\rm L} \Phi_{\rm c} \nu_{\rm R} + {\rm h.c.}
= \left( m_{\nu} + \frac{h_{\nu}}{\sqrt{2}} \sigma \right) \bar{\nu} \nu
\ee
Standard notation is used, with $l_{\rm L}$ representing a left-handed lepton 
doublet, $\nu_{\rm R}$ a right-handed neutrino singlet, and $\Phi_{\rm c}$ the 
charge-conjugated Higgs field.
For small energy and momentum transfers, the Higgs boson exchange between 
neutrinos can be replaced by the contact interaction
\be
H_{\rm I} = - \frac{h_{\nu}^2}{4m_{\sigma}^2} (\bar{\nu} \nu) (\bar{\nu} \nu)
\label{HI}
\ee
with $m_{\sigma}$ the Higgs mass.  This interaction is attractive. It can be 
derived by solving the field equation for $\sigma$ in terms of the neutrino 
field and substituting back into the Lagrangian.  In Dirac representation we 
express the neutrino field as
\ba
\nu_{\rm L} = \thalf (1 - \gamma_5) \nu = \frac{1}{\sqrt{2}}
\left( \begin{array}{c} \psi_{\rm L} \\ -\psi_{\rm L} \end{array} \right)
\nonumber \\
\nu_{\rm R} = \thalf (1 + \gamma_5) \nu = \frac{1}{\sqrt{2}}
\left( \begin{array}{c} \psi_{\rm R} \\ \psi_{\rm R} \end{array} \right)
\ea
where $\psi_{\rm L}$ and $\psi_{\rm R}$ are two-component spinors.  Then the 
interaction can be written as
\be
H_{\rm I} = - \frac{h_{\nu}^2}{ 4m_{\sigma}^2} \left\{
2 \psi_{{\rm L}a}^{\dagger}
\psi_{{\rm R}b}^{\dagger} \psi_{{\rm L}b} \psi_{{\rm R}a}
+ \psi_{{\rm L}a}^{\dagger}
\psi_{{\rm L}b}^{\dagger} \psi_{{\rm R}b} \psi_{{\rm R}a}
+ \psi_{{\rm R}a}^{\dagger}
\psi_{{\rm R}b}^{\dagger} \psi_{{\rm L}b} \psi_{{\rm L}a} \right\}
\ee
where the summation over the spinor indices $a,b$ runs from 1 to 2. Now the 
familiar path to Cooper pairing in the BCS theory follows naturally 
\cite{BCSbooks}.

Allowing for condensation of the form
\be
\langle \psi_{\rm L}^a \psi_{\rm R}^b \rangle = \varepsilon^{ab} D
\ee
where $\varepsilon^{ab}$ is the Levi-Civita symbol and $D$ is related to the 
gap, corresponds to spin-0 pairing of left and right-handed neutrinos 
\cite{ansatz}.  As mentioned earlier, pairing of left-handed neutrinos with 
themselves is not considered because of the repulsive $Z^0$ exchange.  Making 
the mean field approximation results in the 
interaction
\be
H_{\rm I}^{\rm MF} = \frac{h_{\nu}^2}{ 2m_{\sigma}^2}
\left( D \psi_{{\rm L}a}^{\dagger}
\psi_{{\rm R}b}^{\dagger} + D^* \psi_{{\rm L}b} \psi_{{\rm R}a}
\right) \varepsilon^{ab}
\ee
In terms of particle creation and annihilation operators this is \cite{com1}
\ba
H_{\rm I}^{\rm MF} &=& - \frac{h_{\nu}^2}{4 m_{\sigma}^2}
\sum_{\bf p} \frac{m_{\nu}}{\epsilon}
\left\{ D {\rm e}^{2i\epsilon t} \left[ b^{\dagger}(p,+) b^{\dagger}(-p,-) -
b^{\dagger}(p,-) b^{\dagger}(-p,+) \right] \right. \nonumber \\
& & \mbox{} \left. + D^* {\rm e}^{-2i\epsilon t} \left[ b(-p,-) b(p,+)
- b(-p,+) b(p,-) \right] \right\}
\ea
where $\epsilon = \sqrt{p^2+m_{\nu}^2}$. The operator $b^{\dagger}(p,+)$ creates 
a particle with momentum ${\bf p}$ and spin projection $s_z = \thalf$, and so 
on.  To this must be added the free particle Hamiltonian
\be
H_{\rm free} = \sum_{\bf p} \epsilon
\left[ b^{\dagger}(p,+) b(p,+) + b^{\dagger}(p,-) b(p,-) \right]
\ee
The full Hamiltonian, including the chemical potential $\mu$, can be put in the 
standard form
\ba
H &=& \sum_{\bf p} E \left[ c^{\dagger}(p,+) c(p,+) + c^{\dagger}(p,-) c(p,-) 
\right] \\
E &=& \sqrt{ (\epsilon - \mu)^2 + (Km_{\nu}/\epsilon)^2}
\ea
where $K = h_{\nu}^2 |D|/ 2m_{\sigma}^2$, which has units of energy.  This is 
accomplished by the time-dependent canonical transformation
\ba
c(p,+) &=& \cos \theta {\rm e}^{-i(\alpha + \epsilon t)} b(p,+) -
\sin \theta {\rm e}^{i(\alpha + \epsilon t)} b^{\dagger}(-p,-) \nonumber \\
c(p,-) &=& \cos \theta {\rm e}^{-i(\alpha + \epsilon t)} b(p,-) +
\sin \theta {\rm e}^{i(\alpha + \epsilon t)} b^{\dagger}(-p,+)
\ea
with
\be
\tan (2\theta) = \frac{Km_{\nu}}{\epsilon (\epsilon - \mu)}
\ee
Here $D = |D| {\rm e}^{2i\alpha}$.

The gap equation may be derived by demanding self-consistency between the 
assumed value of the condensate and the value obtained from the canonical 
transformation of the creation and annihilation operators.  One finds that 
either $D = 0$ or $\alpha = \pi/2$ with the magnitude of the gap determined by
\be
\frac{h_{\nu}^2}{8 m_{\sigma}^2} \int \frac{d^3p}{(2\pi)^3} 
\frac{m_{\nu}^2}{\epsilon^2}
\frac{1}{\sqrt{(\epsilon - \mu)^2 + (Km_{\nu}/\epsilon)^2}} = 1
\label{basic}
\ee
The integral is divergent.  It could be cut-off with an upper limit $\Lambda$ of 
order $m_{\sigma}$ or with the form-factor
$m_{\sigma}^2/(m_{\sigma}^2 + 4 {\bf p}^2)$ befitting the Higgs boson exchange 
interaction.  The gap equation has a nontrivial solution with $K \neq 0$, no 
matter how small the coupling of the neutrino to the Higgs (expressed as 
$h_{\nu}$ or $m_{\nu}$), on account of the divergence at the energy
$\epsilon = \mu$.  The gap can immediately be inferred to be
\be
\Delta = Km_{\nu}/\mu
\ee
in the weak coupling limit, $m_{\nu}^2 h_{\nu}^2/8m_{\sigma}^2 < 1$.  Notice 
that there is no pairing and as a consequence no superfluidity when the neutrino 
mass is zero. The reason is simple and was also noticed in the context of color 
superconductivity \cite{Alford}:  Chirality and helicity coincide when the 
neutrino is massless, hence a left-handed neutrino and a right-handed neutrino 
with equal but opposite momenta give a spin-1 projection along the relative 
momentum axis and so cannot contribute to a spin-0 condensate.

The gap equation can be written in a form similar to that of ordinary superconductivity
\be
\thalf g N(0) \int_{\xi_{\rm min}}^{\xi_{\rm max}}
\frac{d\xi}{\sqrt{\xi^2+\Delta^2}} = 1
\ee
where $\xi = \epsilon - \mu$ and $g = h_{\nu}^2/4m_{\sigma}^2$ is the four-point coupling from eq. (\ref{HI}). The phase space density at the Fermi surface, $N(0)$, includes the relativistic factor $m_{\nu}^2/\epsilon^2$ from eq. (\ref{basic})
\be
N(0) = \left. \frac{1}{2\pi^2} p^2 \left(\frac{dp}{d\epsilon}\right)
\frac{m_{\nu}^2}{\epsilon^2} \right|_{\epsilon=\mu} = 
\frac{m_{\nu}^2 v_F}{2\pi^2}
\ee
where $v_F$ is the Fermi velocity.  In ordinary superconductivity the integral is cut off by the Debye frequency $\omega_D$, but here the lower limit is minus the Fermi kinetic energy, $\xi_{\rm min} = -(\mu - m_{\nu})$, and the upper limit is essentially the Higgs boson mass, $\xi_{\rm max} = \Lambda = \sqrt{m_{\sigma}^2 + m_{\nu}^2} - \mu$.  The solution to the gap equation is
\ba
\Delta &=& 2 \sqrt{|\xi_{\rm min}| \xi_{\rm max}}
\, \exp[-1/gN(0)] \nonumber \\  
&=& 2 \sqrt{(\mu-m_{\nu})\Lambda}
\, \exp[-8\pi^2 m_{\sigma}^2/h_{\nu}^2 m_{\nu}^2 v_F]
\label{gaps}
\ea
Since this is a typical BCS-like theory, the critical temperature takes the standard form \cite{BCSbooks,Pisarski}
\be
T_c = \frac{{\rm e}^{\gamma}}{\pi} \Delta \approx 0.57 \, \Delta
\ee
When the coupling is not weak the gap equation can be solved numerically.  Up to 
this point, all formulas have been expressed in terms of $h_{\nu}$ and $m_{\nu}$ 
without making the connection $m_{\nu} = h_{\nu} v_0/\sqrt{2}$.

Unfortunately, for the observed set of three flavors of neutrinos the numbers 
are uninterestingly small.  For example, for a neutrino of mass 1 eV and a Higgs 
boson of mass 110 GeV the argument of the exponential in the relativistic limit 
of eq. (\ref{gaps}) is about $-10^{46}$.  Therefore, let us consider a very 
heavy neutrino and the corresponding nonrelativistic limit.  The number of
anti-neutrinos is assumed to be negligible in comparison to the number of neutrinos (or vice versa), just as is the case for baryon number.  In terms of the neutrino mass density $\rho_{\nu} = m_{\nu} n_{\nu}$, with number density 
appropriate to a cold Fermi gas, $n_{\nu} = p_F^3/3\pi^2$, the gap is
\be
\Delta = \sqrt{\frac{2\Lambda}{m_{\nu}}}
\left( \frac{3\pi^2\rho_{\nu}}{m_{\nu}} \right)^{1/3} {\rm e}^{-x}
\ee
and the Pippard coherence length is
\be
\xi = \frac{v_F}{\pi \Delta} =
\frac{1}{\pi \sqrt{2\Lambda m_{\nu}}} {\rm e}^x
\ee
where
\be
x = \frac{4\pi^2 m_{\sigma}^2 v_0^2}
{m_{\nu}^2 (3\pi^2 \rho_{\nu} m_{\nu}^2)^{1/3}}
\ee
First let us apply these results to cosmology.  Given that $\rho_{\nu}$ should 
not exceed the present energy density of the universe, which is about 5 
keV/cm$^3$, means that the critical temperature $T_c$ is less than 1 K by many 
orders of magnitude no matter what the neutrino mass is.  Hence neutrino 
superfluidity does not seem to be relevant to the cosmological expansion of the 
universe.

Next let us apply these results to neutron stars, where heavy neutrinos may 
accumulate due to gravitational attraction.  Choose a reference mass scale of 10 
TeV and a reference energy density of 10 MeV/fm$^3$, which is about 1\%
central density of a neutron star.  Such a modest energy density will not 
significantly alter the structure of the neutron star.  Then
\ba
\Delta &=& 67.2 \left( \frac{\rho_{\nu}}{10\,{\rm MeV/fm}^3} \right)^{1/3}
\left( \frac{10\,{\rm TeV}}{m_{\nu}} \right)^{4/3} {\rm e}^{-x} \;\; {\rm keV} 
\nonumber \\
\xi &=& 5.71\times 10^{-4} \, {\rm e}^x \;\; {\rm fm}
\nonumber \\
x &=& 4.73 \left( \frac{10\,{\rm MeV/fm}^3}{\rho_{\nu}} \right)^{1/3}
\left( \frac{10\,{\rm TeV}}{m_{\nu}} \right)^{8/3}
\ea
To be of interest, the coherence length should be much less than the radius of a 
neutron star, or about 10 km.  In addition, the critical temperature must be less than the interior temperature of the star.  After 1 million years its interior temperature has dropped to about $10^6$ K \cite{BP}.  The quantities $\Delta$, $\xi$ and $T_c$ are strongly dependent on the neutrino mass.  For example, if $m_{\nu} = 8$, 10, 12 TeV then $\xi = 3.0$, 0.065 and 0.01 fm, while $T_c = 0.11 \times 10^6$, $3.9 \times 10^6$, and $19 \times 10^6$ K, respectively.  So neutrino superfluidity in neutron stars could be interesting if there exist Dirac neutrinos with a mass on the order of, or exceeding, 10 TeV.

In conclusion, it has been demonstrated that neutrino superfluidity is a 
possibility if Dirac neutrinos exist with nonzero mass.  If the neutrino coupled 
to a much lighter scalar boson than the Higgs, or if a much heavier neutrino 
exists, then neutrino superfluidity could conceivably be realized in nature. 

\section*{Acknowledgements}

I am grateful to P. J. Ellis, Y.-Z. Qian, K. Rajagopal, I. Shovkovy, and A. 
Steiner for comments on the manuscript.  This work was supported by the US 
Department of Energy under grant DE-FG02-87ER40328.


\begin{thebibliography} {99}
\bibitem{com1} We can safely neglect the contribution of anti-particles because 
pairing is dominated by the attractive interaction between particles very near 
the Fermi surface.
\end{thebibliography}
\end{document}